\documentclass[aps,prx,twocolumn,superscriptaddress,nofootinbib,longbibliography]{revtex4-2}

\usepackage[T1]{fontenc}
\usepackage{lmodern}
\usepackage{amsmath}
\usepackage{amssymb}  
\usepackage{esint}    
\usepackage{graphicx}
\usepackage{xcolor}
\usepackage[colorlinks,linkcolor=blue,citecolor=blue,urlcolor=blue]{hyperref}
\usepackage{orcidlink}

\begin{document}
\title{Drift-induced nonreciprocity in image-polariton-mediated dipole emission}

\author{Christian~Nicolaisen~Hansen\,\orcidlink{0009-0008-2350-7652}}
\email{nicolaisen@mci.sdu.dk}
\affiliation{POLIMA---Center for Polariton-driven Light--Matter Interactions, University of Southern Denmark, Campusvej 55, DK-5230 Odense M, Denmark}

\author{Eduardo~J.~C.~Dias\,\orcidlink{0000-0002-6347-5631}}
\affiliation{POLIMA---Center for Polariton-driven Light--Matter Interactions, University of Southern Denmark, Campusvej 55, DK-5230 Odense M, Denmark}

\author{Christos~Tserkezis\,\orcidlink{0000-0002-2075-9036}}
\affiliation{POLIMA---Center for Polariton-driven Light--Matter Interactions, University of Southern Denmark, Campusvej 55, DK-5230
Odense M, Denmark}
\affiliation{Danish Institute for Advanced Study, University of Southern Denmark, Campusvej 55, DK-5230 Odense M, Denmark}

\author{Joel~D.~Cox\,\orcidlink{0000-0002-5954-6038}}
\email{cox@mci.sdu.dk}
\affiliation{POLIMA---Center for Polariton-driven Light--Matter Interactions, University of Southern Denmark, Campusvej 55, DK-5230 Odense M, Denmark}
\affiliation{Danish Institute for Advanced Study, University of Southern Denmark, Campusvej 55, DK-5230 Odense M, Denmark}

\date{\today}

\begin{abstract}
Electrical current bias offers a magnet-free route to nonreciprocal light propagation at the nanoscale. Here, we theoretically explore drift-induced nonreciprocity in graphene heterostructures containing hexagonal boron nitride (hBN), where graphene plasmons interact with hBN phonon polaritons in the infrared. Both the random-phase approximation and the Boltzmann transport equation are used to characterize the nonlocal optical response of drift-biased graphene. We compare graphene on hBN with an image-polaritonic structure comprising graphene, a thin hBN spacer, and a gold substrate. Carrier drift produces pronounced directional shifts of the polariton dispersion, while coupling to hBN phonon polaritons strongly modifies the nonreciprocal response. In the image-polaritonic structure, the graphene polaritons become up to twice as confined while retaining a comparable Fizeau shift. The nonreciprocal polariton dispersion produces directional near-field emission and nonreciprocal dipole--dipole interactions, the latter quantified through the cross density of optical states, reaching a left--right dissymmetry of up to $\sim$40\%. Our results identify image polaritons in drift-biased graphene as a compact, electrically tunable platform for directional coupling between quantum emitters at the nanoscale.
\end{abstract}

\maketitle

\section{Introduction}

The ever-increasing demand for more compact and integrated photonic technologies~\cite{Koenderink2015_Science348,Wang2019_NatPhotonics14} is driving the search for new ways of controlling light at the nanoscale, where conventional approaches are limited by optical diffraction~\cite{Hohenester2019}. Polaritons---hybrid excitations of light and matter---offer a compelling route beyond this limit~\cite{Basov2020_Nanophotonics10}. In particular, surface polaritons (SPs) bound to material interfaces confine electromagnetic energy to deeply subwavelength volumes and thereby strongly enhance light--matter interactions, with applications ranging from flat optics~\cite{Yu2014_NatMater13,Galiffi2024_NatRevMater9} to nonlinear~\cite{Kauranen2012_NatPhoton6,Lin2025_Engineering45} and quantum~\cite{Bozhevolnyi2017,GonzalezTudela2024_NatRevPhys6} optics.

Nonreciprocal photonics has emerged as a powerful paradigm for controlling SP propagation~\cite{Camley1987_SurfSciRep7,Caloz2018_PhysRevApplied10}. A system is nonreciprocal when the transmission of light between two points changes upon interchanging source and detector, i.e., when Lorentz reciprocity is broken~\cite{Potton2004_RepProgPhys67}. This is distinct from anisotropy: an anisotropic yet reciprocal medium may guide light preferentially along certain axes, but always supports counter-propagating modes with identical dispersion. In linear, time-invariant media, breaking reciprocity requires an external bias that is odd under time reversal~\cite{Caloz2018_PhysRevApplied10}, and enables one-way propagation, isolation, and circulation of electromagnetic waves~\cite{Yu2008_PhysRevLett100,Davoyan2014_NatCommun5,Li2024_NatCommun15}.

Conventionally, nonreciprocity is induced by a static magnetic field, as in magneto-optical and magnetoplasmonic platforms~\cite{Wang2005_OptLett15,Wang2008_PhysRevLett100,Haldane2008_PhysRevLett100,Raghu2008_PhysRevA78,Davoyan2013_PhysRevLett111,Chin2013_NatCommun4,Hu2012_OptLett11,Maksymov2015_Nanomat5}. However, magnetic biasing requires bulky components that are difficult to integrate, while magneto-optical effects in most materials become weak at mid-infrared and optical frequencies~\cite{Caloz2018_PhysRevApplied10, Morgado2020_PhysRevB102}. These limitations have motivated the search for magnet-free alternatives based on spatiotemporal modulation~\cite{Lira2012_PhysRevLett109,Sounas2013_NatCommun4}, optical nonlinearity~\cite{Fan2012_Science335,Shi2015_NatPhoton9}, and electrical current bias~\cite{Bliokh2018_OptLett43}. In the latter case, a DC current---which is odd under time reversal---renders the conductivity both anisotropic and nonreciprocal: the drifting carriers drag the plasmons along with them, so that plasmons co-propagating with the current acquire a longer wavelength than counter-propagating ones~\cite{Borgnia2015_ArXiv,Dong2021_Nature594,Zhao2021_Nature594,Blevins2023_ACSPhoton11}. By analogy with Fizeau's experiment on light in moving water, this effect is known as plasmonic Fizeau drag.

Graphene~\cite{Novoselov2004_Science306,Novoselov2005_Nature438} is a particularly attractive platform for current-induced nonreciprocity: it supports surface plasmon polaritons (SPPs) that combine extreme field confinement with electrical tunability via electrostatic gating~\cite{Koppens2011_NanoLett11,Grigorenko2012_NatPhoton6,Fei2012_Nature487,Chen2012_Nature487,Brar2013_NanoLett13,Low2014_ACSNano8,GarciaDeAbajo2014_ACSPhoton1}, while its low carrier density and high mobility allow for drift velocities that are a sizable fraction of the plasmon phase velocity~\cite{Dong2021_Nature594,Zhao2021_Nature594}. This produces strongly asymmetric and even near-unidirectional plasmon propagation~\cite{Sabbaghi2015_PhysRevB92,VanDuppen2016_2DMater3,Morgado2018_ACSPhoton2018,Wenger2018_PhysRevB97,Morgado2020_PhysRevB102,Papaj2020_PhysRevLett125,Prudencio2021_Plasmonics16,Morgado2022_Nanophotonics11}, and the resulting Fizeau drag has recently been observed by near-field imaging of current-carrying graphene~\cite{Dong2021_Nature594,Zhao2021_Nature594}.

Introducing a metallic interface near a graphene layer further enriches this platform by enabling the formation of image polaritons (IPs), hybrid modes that arise from the interaction between a polarizable material and its electromagnetic image in a conductor~\cite{Menabde2022_Nanophotonics11}. Because the field barely penetrates the metal, IPs are far more confined than their unscreened counterparts and exhibit a modified, nearly linear dispersion, leading to enhanced light--matter interactions and pronounced near-field effects~\cite{AlonsoGonzalez2017_NatNanotech12,Lundeberg2017_Science357,Iranzo2018_Science360,RodriguezEcharri2019_Optica6}. The addition of a drift current to this geometry has been predicted to cause focusing of the IPs in the direction of the drifting electrons over a large frequency range~\cite{Sammon2021_PhysRevB104}. In realistic devices, however, the spacer is typically hexagonal boron nitride (hBN), a natural hyperbolic material which hosts mid-infrared phonon polaritons that hybridize strongly with graphene plasmons~\cite{Brar2014_NanoLett14,Dai2015_NatNanotech10,Woessner2015_NatureMater14}. How this hybridization shapes drift-induced nonreciprocity, and how it is in turn modified by metallic screening, has so far received little attention.

In this work, we theoretically explore drift-induced nonreciprocity in graphene--hBN heterostructures with and without a metallic substrate, and its consequences for the near-field emission and mutual interaction of dipole emitters. We describe the drift-biased graphene within the random-phase approximation (RPA), which captures nonlocal and interband effects, and compare it with a semiclassical Boltzmann-transport description. For graphene on hBN, we find that coupling to the hBN phonon polaritons quenches the polariton-mediated dipole--dipole interaction, and with it the nonreciprocity, within a spectral band just above the upper phonon resonance.
In the IP structure, the polaritons become more strongly confined, by up to a factor of two near the upper Reststrahlen band, while retaining a comparable Fizeau shift. We probe these effects with dipole emitters, which serve as models of localized near-field sources such as nanotip scatterers and quantum emitters~\cite{Nikitin2016_NaturePhoton10,Menabde2021_NatureCommun12}, and are particularly suited to probing nonreciprocal environments, where drift-induced asymmetries can enable preferential and more efficient transport of optical energy in a single direction~\cite{RodriguezEcharri2026_NatCommun17}. The nonreciprocity is quantified through the cross density of optical states (CDOS) between two emitters, which acquires a dissymmetry of up to $\sim$40\% under drift at a spatial scale compressed by the image hybridization. These results are relevant for directional photon exchange and current-controlled entanglement between quantum emitters~\cite{Lodahl2017_Nature541,Berres2023_OptExpress31}.

\section{Conductivity of drift-biased graphene}

We describe the optical response of current-carrying graphene using two complementary models: a quantum-mechanical description based on the random-phase approximation (RPA), which includes nonlocal intraband and interband transitions, and a semiclassical description based on the Boltzmann transport equation (BTE), restricted to intraband processes.

A DC bias applied along a graphene layer drives a current in which carriers drift with average velocity $\mathbf{v}_\mathrm d=\beta v_\mathrm{F}\hat{\mathbf{x}}$, where $0\leq \beta <1$ and $v_\mathrm{F}\approx c/300$ is the Fermi velocity. In this nonequilibrium steady state, the Fermi--Dirac distribution is skewed along the drift direction~\cite{VanDuppen2016_2DMater3,Morgado2020_PhysRevB102,Blevins2023_ACSPhoton11} according to
\begin{equation}
    f_T(\varepsilon_{s,k}) = \left[\mathrm{e}^{(\varepsilon_{s,k}-\mu-\hbar\mathbf{v}_{\rm d}\cdot\mathbf{k})/k_\mathrm{B}T}+1\right]^{-1} ,
\end{equation}
where $\mu$ is the chemical potential, $\mathbf{k}$ is the electron wave vector, and $\varepsilon_{s,k}=s\hbar v_\mathrm{F}|\mathbf{k}|$ is the energy of the conical conduction ($s=+1$) or valence ($s=-1$) band of graphene. Due to charge conservation, skewing of the electron distribution lowers the Fermi energy from its value $E_\mathrm{F}$ in the absence of drift to
\begin{equation}
    \tilde{E}_{\rm F} = (1-\beta^2)^{3/4} E_\mathrm{F}.
\end{equation}
Imposing this constraint isolates genuine drift effects from those caused by a change in carrier density. In what follows, we adopt the zero-temperature limit, \mbox{$T\to 0$}, such that $\mu\to\tilde{E}_{\rm F}$, and the distribution becomes
\begin{equation}
    f(\varepsilon_{s,k})=\Theta(\tilde{E}_{\rm F}-\varepsilon_{s,k}+\hbar \mathbf{v}_{\rm d}\cdot \mathbf{k}) ,
\end{equation}
where $\Theta$ is the Heaviside step function, corresponding to a Fermi surface deformed into an ellipse of eccentricity $\beta$ and elongated along the drift direction.

Within the RPA, the collisionless (no phenomenological damping) and non-interacting polarization bubble in graphene is written as a density--density response~\cite{Wunsch2006_NewJPhys8,Hwang2007_PhysRevB75,VanDuppen2016_2DMater3}
\begin{align}
  \Pi^0&(\mathbf{Q},\omega) = g \sum_{s,s'=\pm} \int \frac{{\rm d}^2 \mathbf{k}}{(2\pi)^2} \notag\\
  &\times \frac{f(\varepsilon_{s,k})-f(\varepsilon_{s',|\mathbf{k}+\mathbf{Q}|})}{\hbar\omega+\varepsilon_{s,k}-\varepsilon_{s',|\mathbf{k}+\mathbf{Q}|}+\mathrm{i}0^+} F_{ss'}(\mathbf{k},\mathbf{k}+\mathbf{Q}) ,
  \label{eq:Psi0}
\end{align}
where the spin--valley degeneracy is $g=4$, and we define
\begin{equation}
    F_{ss'}(\mathbf{k},\mathbf{k}+\mathbf{Q})=\frac{1}{2}\left(1+ss'\cos \theta_{\mathbf{k},\mathbf{k}+\mathbf{Q}}\right)
\end{equation}
as the band-overlap factor, with $\theta_{\mathbf{k},\mathbf{k}+\mathbf{Q}}$ denoting the angle between vectors $\mathbf{k}$ and $\mathbf{k}+\mathbf{Q}$. Because the drifting distribution is anisotropic, $\Pi^0$ depends on both the magnitude and the direction of the in-plane wave vector, expressed in polar coordinates as $\mathbf{Q}=(Q,\varphi_\mathbf{Q})$. To include a finite phenomenological relaxation time $\tau$ while conserving the local number of particles, we follow Mermin's prescription \cite{Mermin1970_PhysRevB1} to write the polarization of graphene as
\begin{equation}
    \Pi(\mathbf{Q},\omega)=\frac{(1+\mathrm{i}/\omega\tau)\Pi^0(\mathbf{Q},\omega+\mathrm{i}/\tau)}{1+(\mathrm{i}/\omega\tau)\Pi^0(\mathbf{Q},\omega+\mathrm{i}/\tau)/\Pi^0(\mathbf{Q},0)} .
\end{equation}
Through the continuity equation, $\Pi$ determines the longitudinal optical conductivity, i.e., the projection of the conductivity tensor onto $\hat{\mathbf{Q}}$,
\begin{equation}
    \sigma^\mathrm{RPA}(\mathbf{Q},\omega)=\frac{\mathrm{i}e^2\omega}{Q^2}\Pi(\mathbf{Q},\omega) .
\end{equation}

For comparison, we also consider the semiclassical conductivity of drift-biased graphene obtained from the BTE in the highly doped, intraband-only regime~\cite{Blevins2023_ACSPhoton11,RodriguezEcharri2026_NatCommun17}. In that approach, one linearizes the BTE around a skewed Fermi--Dirac distribution and extracts the conductivity tensor from Ohm's law, thereby neglecting the interband contribution that is naturally present in the full RPA response. This approach allows the extraction of the full tensorial nature of the conductivity, where we, in this work, only require the longitudinal optical conductivity in the reference frame of the incident electromagnetic wave~\cite{Kortkamp2015_PhysRevB92,Abrantes2023_PhysRevB108}, which for a finite drift current along $\hat{\mathbf{x}}$ takes the form
\begin{align}
    \sigma^\mathrm{BTE}(\mathbf{Q},\omega) = \frac{\sigma_\omega}{\pi}&\int_0^{2\pi}{\rm d}\varphi\ \frac{\cos(\varphi-\varphi_\mathbf{Q})}{1+\alpha \cos(\varphi-\varphi_\mathbf{Q})}\notag\\ &\times \frac{\cos(\varphi-\varphi_\mathbf{Q}) - \beta \cos\varphi_\mathbf{Q}}{(1-\beta\cos\varphi)^2}.
\end{align}
Here, we have defined \mbox{$\alpha \equiv -{v_\mathrm{F}Q}/({\omega+\mathrm{i}/\tau})$}, and
\begin{equation}
    \sigma_\omega= \frac{\mathrm{i}e^2}{\pi\hbar^2}\frac{\tilde{E}_{\mathrm F}}{\omega+\mathrm{i}/\tau}
\end{equation}
is the local Drude conductivity evaluated at the reduced Fermi energy $\tilde{E}_{\rm F}$. The BTE description is expected to hold for $\hbar\omega\ll 2E_\mathrm{F}$ and $Q\ll k_\mathrm{F}=E_\mathrm{F}/\hbar v_\mathrm{F}$; beyond this regime, interband transitions and quantum nonlocal corrections captured by the RPA become significant.

\section{Plasmons in current-biased graphene heterostructures}

\begin{figure*}[htbp]
\centering
\includegraphics[width=0.99\linewidth]{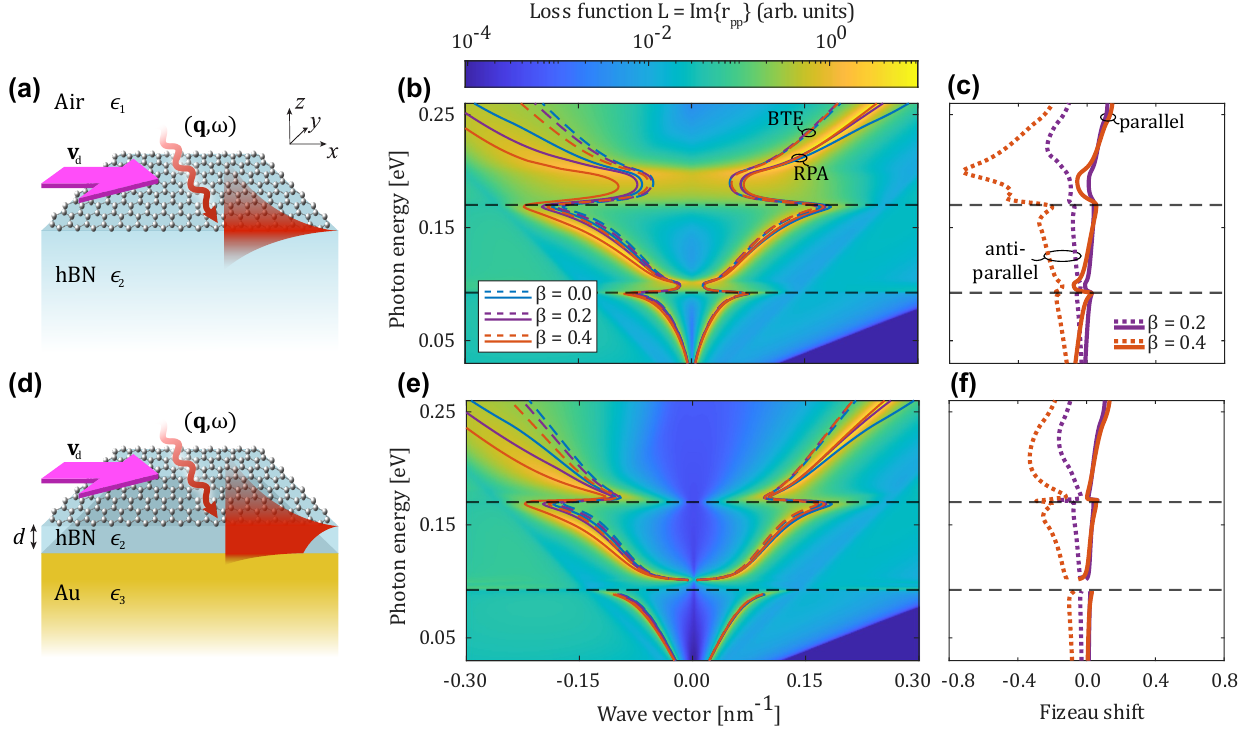}
\caption{\textbf{Loss function and Fizeau shift of drift-biased graphene heterostructures.}
(a) Schematic of the graphene-on-hBN structure, consisting of a graphene layer placed on a semi-infinite hBN substrate of relative permittivity $\epsilon_2$, with a drift current $\mathbf{v}_\mathrm d$ along the graphene (purple arrow), and incident electromagnetic radiation with in-plane wave vector $\mathbf{Q}$ and frequency $\omega$ (red arrow).
(b) Loss function of the graphene-on-hBN structure calculated using the RPA conductivity for a drift current $\beta=0.4$, where positive (negative) wave vectors correspond to incident light (anti-) parallel to the drift current. The solid (dashed) curves show the dispersion calculated using the RPA (BTE) conductivity for the drift currents indicated in the legend.
(c) Fizeau shift of the graphene-on-hBN structure obtained using the RPA conductivity, with solid (dotted) lines corresponding to the (anti-) parallel direction.
(d) Schematic of the IP structure, consisting of a graphene layer placed on an hBN spacer of relative permittivity $\epsilon_2$ and thickness $d$, on top of a semi-infinite Au substrate of relative permittivity $\epsilon_3$.
(e) Loss function and (f) Fizeau shift of the IP structure with $d=5$\,nm, as in (b) and (c), respectively. In all panels, $E_\mathrm{F}=0.2$\,eV, and the dashed horizontal lines indicate hBN phonon resonances.
}
\label{fig:RPA_disp}
\end{figure*}

We now consider situations in which the drift-biased graphene layer is embedded in two layered, non-magnetic heterostructures stacked along $\hat{\mathbf{z}}$ and infinitely extended in the $xy$ plane, with air ($\epsilon_1=1$) as the superstrate. In the first case, graphene is placed on a semi-infinite hBN substrate of relative permittivity $\epsilon_2$ [Fig.~\ref{fig:RPA_disp}\textcolor{blue}{a}]. In the second, graphene is separated from a semi-infinite gold (Au) substrate of relative permittivity $\epsilon_3$ by an hBN spacer of thickness $d$ [Fig.~\ref{fig:RPA_disp}\textcolor{blue}{d}]. We refer to the latter as the IP structure, since the graphene plasmons couple to their mirror image in the metal to form image polaritons. Optical excitations incident from the superstrate are characterized by an in-plane wave vector $\mathbf{Q}$ and an out-of-plane component $q_{1z}$, such that the full wave vector is $\mathbf{q}_1=(\mathbf{Q},q_{1z})$, and we assume a harmonic time dependence $\mathrm{e}^{-\mathrm{i}\omega t}$.

hBN is a uniaxial crystal with its optical axis along $z$, characterized by in-plane and out-of-plane permittivities $\epsilon_2^{xy}$ and $\epsilon_2^{z}$, for which we adopt the ``damaged'' model given in the supplementary information of Ref.~\cite{Woessner2015_NatureMater14}, based on the first-principles calculations of Ref.~\cite{Cai2007_SolidStateCommun141}. In the mid-infrared, hBN supports phonon polaritons arising from the coupling of light to optical lattice vibrations. Because the in-plane and out-of-plane phonons oscillate at different frequencies, hBN exhibits two Reststrahlen bands in which $\epsilon_2^{xy}$ and $\epsilon_2^{z}$ have opposite signs. Within these bands, the isofrequency surfaces are hyperbolic and hBN supports phonon polaritons with very large in-plane wave vectors~\cite{Caldwell2014_NatCommun5,Dai2015_NatNanotech10}, which hybridize with graphene plasmons and, in the IP structure, are additionally screened by the metal. The Au substrate is described by the Drude model with $\hbar\omega_\mathrm p = 8.84$\,eV, $\hbar\gamma_\mathrm m=103$\,meV, and $\epsilon_\infty=9.84$~\cite{Dias2018_PhysRevB97}.

To investigate the manifestation of nonreciprocity in the dispersion of the heterostructures, we consider monolayer graphene with a Fermi energy of $E_\mathrm{F}= 0.2$\,eV. It has been shown that lower Fermi energies generate the largest Fizeau shifts~\cite{Dong2021_Nature594}, and at this value the graphene plasmon dispersion spans both Reststrahlen bands of hBN, allowing for interactions between graphene plasmons and phonon polaritons. At these low energies, graphene plasmons lie in the mid-infrared and are so highly confined that they have a very large momentum mismatch with freely propagating photons, which makes their excitation from the far field experimentally challenging. Instead, the plasmonic modes can be excited and detected using scattering-type scanning near-field optical microscopy (s-SNOM)~\cite{Fei2012_Nature487,Chen2012_Nature487}. We consider a DC electron mobility of $\mu^\mathrm{DC} = 10{,}000\,\mathrm{cm^2V^{-1}s^{-1}}$ throughout this work, giving a carrier relaxation time~\cite{GarciaDeAbajo2014_ACSPhoton1} of $\tau = \mu^\mathrm{DC}E_\mathrm{F}/ev_\mathrm{F}^2 \approx 0.2\,\mathrm{ps}$. Since the plasmon wave vectors greatly exceed the free-space wave vector ($Q\gg\omega/c$), retardation effects can be safely neglected, allowing the reflection coefficients to be expressed in terms of the electrostatic response of the heterostructures. The reflection coefficients for the graphene-on-hBN structure of Fig.~\ref{fig:RPA_disp}\textcolor{blue}{a} and the IP structure of Fig.~\ref{fig:RPA_disp}\textcolor{blue}{d} both take the form
\begin{align}
    r_\mathrm{pp}&= -\frac{\epsilon_1 - \epsilon_2\xi(Q) - \mathrm{i}Q\sigma(\mathbf{Q},\omega)/\omega\epsilon_0}{\epsilon_1 + \epsilon_2\xi(Q) + \mathrm{i}Q\sigma(\mathbf{Q},\omega)/\omega\epsilon_0} ,
\end{align}
where $\epsilon_{0}$ is the vacuum permittivity, and we define \mbox{$\epsilon_2\equiv \sqrt{\epsilon_2^{xy} \epsilon_2^z}$}. The form of $\xi(Q)$ depends on the given heterostructure, taking the value $\xi(Q) = 1$ for the graphene-on-hBN structure and
\begin{equation}
    \xi(Q) = \frac{\epsilon_3 + \epsilon_2\tanh(\bar \epsilon_2 Q d)}{\epsilon_2 + \epsilon_3\tanh(\bar \epsilon_2 Q d)} ,
\end{equation}
where $ \bar \epsilon_2\equiv \sqrt{\epsilon_2^{xy} / \epsilon_2^z}$, for the IP structure. Since the quasistatic electric field is predominantly longitudinal, the graphene layer enters $r_\mathrm{pp}$ only through its longitudinal conductivity $\sigma(\mathbf{Q},\omega)$. For the IP structure, the spacer has a finite thickness chosen at $d=5$\,nm, which corresponds to approximately 15 atomic monolayers of hBN~\cite{Iranzo2018_Science360}.

The contour map of Fig.~\ref{fig:RPA_disp}\textcolor{blue}{b} shows the loss function $L=\mathrm{Im}\{r_\mathrm{pp}\}$ of the graphene-on-hBN structure, where the graphene conductivity is calculated using the RPA with a drift current of $\beta=0.4$. The positive (negative) values of the in-plane wave vector $Q$ describe incident radiation with an in-plane component (anti-) parallel to the direction of the drift current. The loss maxima indicate the dispersion curves of the system. At energies above $\sim0.2$\,eV, the maximum is much more pronounced in the parallel direction (right side of the map) than in the anti-parallel direction (left side), where it is diffuse; this nonreciprocity appears due to the applied drift current. For reference, the dispersion curve of the structure without applying a drift current ($\beta=0$) is inserted as the blue curves; as expected, these are identical for the anti-parallel and parallel directions. The dispersion lines of finite drift currents $\beta=0.2$ and $\beta=0.4$ are inserted as the purple and red curves, respectively. These dispersion curves are calculated using RPA (BTE) and shown as the solid (dashed) curves. Especially at larger energies, the effect of increasing the drift current is to shift the dispersion to lower wave vectors in the parallel direction and to higher wave vectors in the anti-parallel direction. The interband transitions included within the RPA strongly affect the nonreciprocity of the dispersion at frequencies near and above the Fermi energy. The RPA dispersion curves differ substantially from those predicted by the BTE, and they also vary more strongly with drift current. In what follows, we focus our discussion on the RPA conductivity.

The phonon resonances of the hBN layer are indicated by the horizontal dashed black lines. Hybridization of phonon polaritons with graphene plasmons causes an anticrossing, manifesting as a splitting in the loss function peaks and back-bending of the dispersion curves just above both phonon resonances~\cite{Brar2014_NanoLett14,Dai2015_NatNanotech10,Woessner2015_NatureMater14}.

The Fizeau shift of the graphene-on-hBN structure, obtained from the RPA dispersion curves for finite drift currents $\beta=0.2$ and $\beta=0.4$, is shown in Fig.~\ref{fig:RPA_disp}\textcolor{blue}{c}. For a given frequency $\omega$ and angle $\varphi_\mathbf{Q}$, the Fizeau shift $\Delta\lambda_\mathrm p/\lambda_\mathrm p$ is defined as the change $\Delta\lambda_\mathrm p$ in the plasmon wavelength induced by the drift current, normalized to the undrifted plasmon wavelength $\lambda_\mathrm p$. The plasmon wavelengths are computed from the reflection coefficient pole $Q_\mathrm p$ as $\lambda_\mathrm p= 2\pi /Q_\mathrm p$ for a given drift current. For the anti-parallel direction, the Fizeau shift typically takes a negative value, while for the parallel direction it is mostly positive and much smaller in magnitude. As expected, the magnitude of the shift increases for a larger drift current in the anti-parallel direction, but the shift is largely unaffected by a larger drift current in the parallel direction. Around the phonon resonances, the Fizeau shifts decrease in magnitude as the hybridized polaritons become more phononic in nature, and the drift current becomes less influential on the resulting dispersion curves.

\begin{figure*}[t!]
\centering
\includegraphics[width=0.99\linewidth]{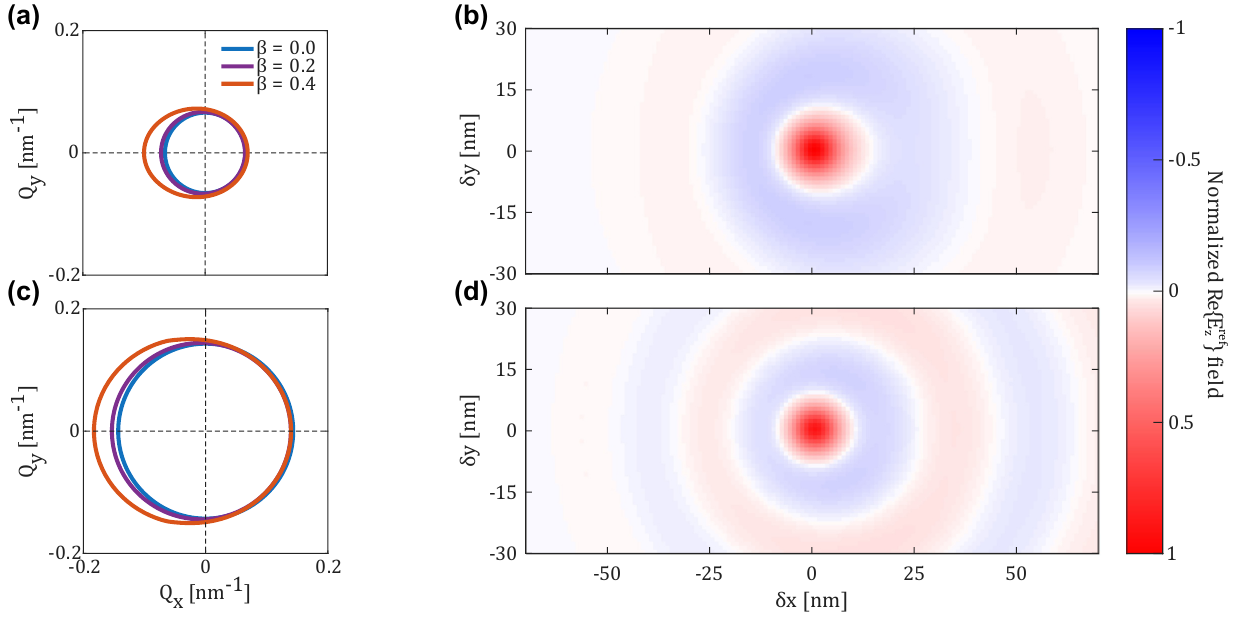}
\caption{\textbf{Isofrequency contours and reflected field of an electric dipole placed above drift-biased graphene heterostructures.}
(a) Isofrequency contours of the graphene-on-hBN structure for the drift currents indicated in the legend.
(b) Real part of the reflected out-of-plane field component, $\mathrm{Re}\{E_z^\mathrm{ref}\}$, produced by an electric dipole $\hat{\mathbf{p}}_1 = \hat{\mathbf{z}}$ placed at $z=4$\,nm above the graphene-on-hBN structure for $\beta=0.4$.
(c) and (d) Same as (a) and (b), respectively, for the IP structure with $d=5$\,nm.
In all panels, $E_\mathrm{F}=0.2$\,eV and $\hbar\omega= 0.2$\,eV.
}
\label{fig:Map_beta03}
\end{figure*}

Moving on to the IP structure sketched in Fig.~\ref{fig:RPA_disp}\textcolor{blue}{d}, the corresponding loss function is shown in Fig.~\ref{fig:RPA_disp}\textcolor{blue}{e}. The interaction of the graphene plasmons with their image from the interface with the Au substrate causes a shift in the dispersion curves towards higher wave vectors. Just above the phonon resonances, where for the graphene-on-hBN structure we had a continuum of allowed phonon modes, the finite thickness now restricts the allowed modes. For the lower phonon resonance, this causes an energy gap to appear in the dispersion curves, and for the upper phonon resonance, the back-bending appears across a much narrower energy band.

While the dispersion curves for all drift currents are shifted to higher wave vectors in both the parallel and anti-parallel directions due to the image interaction, the relative shifts in the dispersion curves from $\beta=0$ to finite drift currents $\beta=0.2$ and $\beta=0.4$ are slightly reduced compared to the graphene-on-hBN structure, as indicated by the decrease in the Fizeau shifts shown in Fig.~\ref{fig:RPA_disp}\textcolor{blue}{f}. Exceptions to this appear just above the phonon resonances: in the energy gap of the lower phonon resonance, the Fizeau shift is undefined since no mode exists, while at the upper phonon resonance the Fizeau shift in the parallel direction is strongly reduced.

\section{Nonreciprocal dipole interactions}

\begin{figure*}[t!]
\centering
\includegraphics[width=0.99\linewidth]{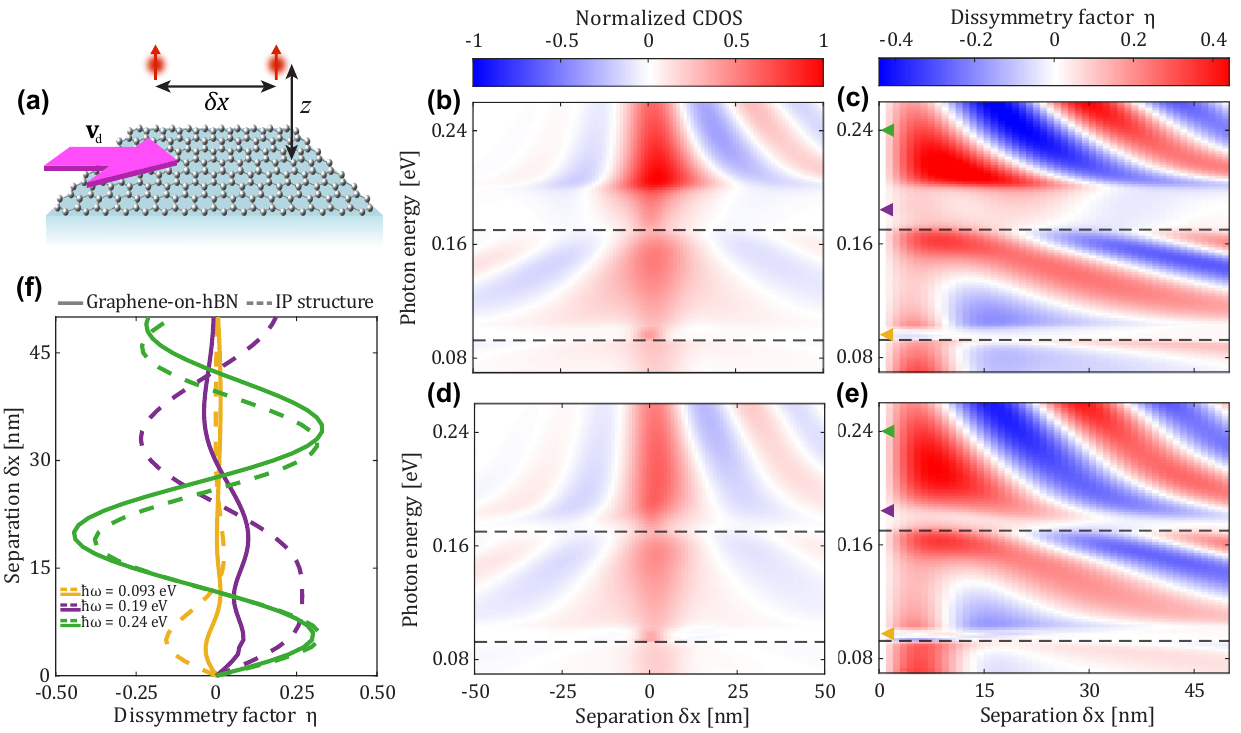}
\caption{\textbf{CDOS and dissymmetry factor of drift-biased graphene heterostructures.}
(a) Sketch of two dipoles placed a distance $z$ above either of the graphene heterostructures and separated by a distance $\delta x$ along the drift direction.
(b) Normalized CDOS and (c) dissymmetry factor for two dipoles placed above the graphene-on-hBN structure.
(d) Normalized CDOS and (e) dissymmetry factor for two dipoles placed above the IP structure with $d=5$\,nm.
(f) Dissymmetry factor for the graphene-on-hBN (IP) structure, shown as solid (dashed) curves, at the photon energies indicated by the colored arrows shown in (c) and (e).
In all panels, $E_\mathrm{F}=0.2$\,eV, $\beta=0.4$, and both dipoles are placed at $z=4$\,nm above the graphene and oriented along $\hat{\mathbf{p}}_{1,2}=\hat{\mathbf{z}}$. The dashed horizontal lines indicate hBN phonon resonances.
}
\label{fig:CDOS_beta03}
\end{figure*}

We consider a light scatterer oscillating at photon energy $\hbar\omega= 0.2\ \mathrm{eV}$, such as a nanoparticle or quantum emitter, modeled as an electric dipole with dipole moment $\mathbf{p}_1 = \hat{\mathbf{z}}$ placed at a distance $z_1=4$\,nm above the heterostructures. A dipole oriented normal to the interface also provides a useful first approximation to the tip of an s-SNOM setup: the metallic tip is typically elongated along the direction normal to the sample surface, and illumination polarized along the tip axis therefore produces a dominant dipole moment along this direction~\cite{Nikitin2016_NaturePhoton10,Menabde2021_NatureCommun12}. The electric field at a point $\mathbf{r}_2$ produced by the light scatterer at $\mathbf{r}_1$ and reflected by the drift-biased graphene heterostructure is given as
\begin{equation}
    \mathbf{E}^\mathrm{ref}(\mathbf{r}_2) = \frac{\omega^2}{\epsilon_0 c^2} \mathbf{G}^\mathrm{ref}(\mathbf r_2,\mathbf r_1;\omega)\cdot \mathbf{p}_1,
\end{equation}
where, within the plane $z_1=z_2\equiv z$ above the graphene, the dyadic Green's tensor element that yields the $z$ component of the reflected electric field for a dipole oriented along $\hat{\mathbf{z}}$ is given in the quasistatic limit as
\begin{align}
    G^\mathrm{ref}_{33} & (\mathbf{r}_2,\mathbf{r}_1;\omega)=\frac{1}{8 \pi^2 q_1^2}\int_0^\infty \mathrm{d}Q \ Q^2\mathrm{e}^{-2  Qz} \notag\\
    &\times \int _0^{2\pi}\mathrm d\varphi_\mathbf{Q}\,r_\mathrm{pp}\,\mathrm{e}^{\mathrm{i}Q(\delta x\cos \varphi_\mathbf{Q} + \delta y\sin \varphi_\mathbf{Q})} . \label{Eq:Gref33}
\end{align}
Here, $q_1=\sqrt{\epsilon_1}\,\omega/c$ is the magnitude of the wave vector in the superstrate, while $\delta x = x_2-x_1$ and $\delta y = y_2-y_1$ denote the in-plane distances from the scatterer to the observation point.

At $\hbar\omega= 0.2$\,eV, where the dispersion differs strongly between the parallel and anti-parallel directions [Fig.~\ref{fig:RPA_disp}\textcolor{blue}{b} and \textcolor{blue}{e}], the isofrequency contour of the polaritons supported by the graphene-on-hBN structure is shown in Fig.~\ref{fig:Map_beta03}\textcolor{blue}{a}. With no drift bias, the contour is perfectly circular (blue curve), given that the graphene layer is isotropic. As the drift current is increased to $\beta = 0.2$, the symmetry is broken, with the supported polaritons propagating anti-parallel to the drift current becoming slightly more confined and lossy, an effect that is further enhanced as the drift current increases to $\beta = 0.4$. This asymmetry for $\beta=0.4$ is seen in the real $z$ component of the reflected field from the electric dipole, $\mathrm{Re}\{E_z^\mathrm{ref}\}$, in Fig.~\ref{fig:Map_beta03}\textcolor{blue}{b} for the graphene-on-hBN structure, where we observe partial unidirectionality in the emission to either side of the dipole along $x$. The reflected fields decay very quickly because, at this energy just above the upper phonon resonance, the plasmon--phonon interaction provides a large loss channel.

For the IP structure, Fig.~\ref{fig:Map_beta03}\textcolor{blue}{c} shows the polariton isofrequency contour, which reaches larger wave vectors due to the image interaction, and we again see a broken symmetry with increasing drift current. Fig.~\ref{fig:Map_beta03}\textcolor{blue}{d} shows $\mathrm{Re}\{E_z^\mathrm{ref}\}$ from the electric dipole above the IP structure, where the restricted set of phonon modes in the finite thickness hBN layer leads to a slower field decay, while still showing a degree of unidirectionality along $x$. The larger wave vectors of the supported polaritons lead to ripples in the reflected electric field that are compressed to smaller distances as compared to the graphene-on-hBN structure. Incidentally, nonreciprocity can also be observed in the fields emitted by electric dipoles oriented parallel to the graphene plane (see Supplemental Material).

We now consider the interaction between two light scatterers placed above the drift-biased graphene heterostructures. This dipole--dipole interaction is quantified by the CDOS, which for electric dipoles $\mathbf{p}_1$ and $\mathbf{p}_2$ placed at $\mathbf{r}_1$ and $\mathbf{r}_2$, respectively, is given by~\cite{Caze2013_PhysRevLett110,Hohenester2019}
\begin{equation}
    \rho_{12}(\omega) = \frac{2\omega}{\pi c^2}\mathrm{Im}\{\hat{\mathbf{p}}_2^\ast \cdot \mathbf G^\mathrm{tot}(\mathbf r_2,\mathbf r_1;\omega) \cdot \hat{\mathbf{p}}_1\} ,
\end{equation}
where $\hat{\mathbf{p}}_i$ are the unit vectors that point along the electric dipole moments, and $\mathbf G^\mathrm{tot} = \mathbf G^0 + \mathbf G^\mathrm{ref}$ contains both the homogeneous-space solution $\mathbf G^0$ and the Green's dyadic due to the reflection at the heterostructure $\mathbf G^\mathrm{ref}$. In a reciprocal environment, $\mathbf G^\mathrm{tot}(\mathbf{r}_2,\mathbf{r}_1;\omega)=[\mathbf G^\mathrm{tot}(\mathbf{r}_1,\mathbf{r}_2;\omega)]^{\rm T}$, so that the CDOS is unchanged when the roles of the two dipoles are interchanged. The drift current breaks this symmetry, rendering the dipole--dipole interaction itself nonreciprocal.

For both dipoles pointing in the out-of-plane direction $\hat{\mathbf{p}}_i =\hat{\mathbf{z}}$, the CDOS is computed from a single Green's dyadic tensor element $G^\mathrm{tot}_{33}(\mathbf r_2,\mathbf r_1;\omega)$, where the homogeneous term can be found in standard textbooks~\cite{Novotny2025}. As for the reflected electric field, we are interested in the CDOS along the drift current direction $x$, and we thus consider two dipoles at the same height $z_1=z_2\equiv z$ with $\delta y=0$, and allow their separation $\delta x = x_2-x_1$ to vary, as shown in Fig.~\ref{fig:CDOS_beta03}\textcolor{blue}{a}. The reflected term of the Green's dyadic tensor element for this setup is again determined by Eq.~\eqref{Eq:Gref33}. Figs.~\ref{fig:CDOS_beta03}\textcolor{blue}{b} and \ref{fig:CDOS_beta03}\textcolor{blue}{d} show the CDOS for graphene-on-hBN and the IP structure, respectively; for both structures, the nonreciprocity is seen as an asymmetry in the CDOS fringes around $\delta x=0$. For graphene-on-hBN, there is an energy band just above the upper phonon resonances (horizontal dashed line) where the CDOS is close to zero away from $\delta x=0$, as the interaction of nonreciprocal graphene plasmons with phonon polaritons in the hBN layer reduces the reflected fields. In contrast, for the IP structure, where the allowed phonon modes are restricted by the finite thickness of the hBN layer, the CDOS fringes extend through these energy bands.

To quantify the nonreciprocity of the system along the drift current direction, we define a dissymmetry factor
\begin{equation}
    \eta = \left.\frac{\rho_{12} - \rho_{21}}{\rho_{11}}\right|_{\delta y=0} ,
\end{equation}
as the difference between the CDOS $\rho_{12}(\delta x>0)$ for $\mathbf{p}_2$ placed to the right of $\mathbf{p}_1$ and the CDOS \mbox{$\rho_{21}(\delta x) = \rho_{12}(-\delta x)$} for $\mathbf{p}_2$ placed at the same distance to the left of $\mathbf{p}_1$, normalized to the local density of optical states $\rho_{11}$, i.e., the value of $\rho_{12}$ at $\delta x=0$. Fig.~\ref{fig:CDOS_beta03}\textcolor{blue}{c} shows the dissymmetry factor for graphene-on-hBN, where the band of low CDOS above the upper phonon resonance appears as a low dissymmetry factor at larger separations, and Fig.~\ref{fig:CDOS_beta03}\textcolor{blue}{e} shows the dissymmetry factor for the IP structure. Away from the phonon resonances, both structures reach similar values of $|\eta|$, up to $\sim0.4$. The dissymmetry factor at the photon energies marked by the colored arrows ($\hbar\omega=0.093$, $0.19$, and $0.24$\,eV) is compared for both heterostructures in Fig.~\ref{fig:CDOS_beta03}\textcolor{blue}{f}. At $0.24$\,eV (green), the two structures yield nearly identical dissymmetry, whereas at $0.19$\,eV (purple), within the band above the upper phonon resonance, $|\eta|$ remains below $\sim0.1$ for graphene-on-hBN but reaches $\sim0.27$ for the IP structure. Near the lower phonon resonance ($0.093$\,eV, yellow), the dissymmetry is weak for both structures.

\section{Conclusion}

We have shown that metallic screening offers a route to compress drift-biased graphene polaritons without sacrificing their nonreciprocity. In a graphene--hBN--Au image-polaritonic structure, the polaritons are up to twice as confined as for graphene on bulk hBN, yet they retain a comparable Fizeau shift over most of the mid-infrared range considered.

The interaction between graphene plasmons and the phonon polaritons of hBN substantially modifies the nonreciprocal response. On a bulk hBN substrate, the continuum of phonon modes just above the upper phonon resonance acts as a strong loss channel, quenching the polariton-mediated dipole--dipole interaction, and with it the nonreciprocity, within this spectral band. In the IP structure, the finite spacer thickness restricts the available phonon modes, so that the nonreciprocal response persists across this band. We further find that a full RPA treatment of the drifting electrons is essential: at the large wave vectors and photon energies near the Fermi energy relevant here, interband transitions and quantum nonlocal effects, both absent in the semiclassical Boltzmann description, substantially alter the drift-induced shifts of the dispersion.

We also demonstrated that the nonreciprocal dispersion is directly manifested in the near-field response of localized emitters. A dipole positioned above the drift-biased heterostructures produces asymmetric reflected fields and partially unidirectional propagation along the drift axis, while the cross density of optical states between the two dipoles separated along this axis acquires a sizable dissymmetry ($\sim0.4$ for the parameters considered here). The IP structure retains a comparable degree of nonreciprocity, and preserves it within the band where it is quenched for graphene on bulk hBN. Since plasmonic Fizeau drag has already been imaged in current-carrying graphene by near field microscopy~\cite{Dong2021_Nature594,Zhao2021_Nature594}, these effects could be explored experimentally, offering an electrically tunable route to directional photon exchange and current-controlled entanglement between quantum emitters~\cite{Lodahl2017_Nature541,Berres2023_OptExpress31}.

\section*{Acknowledgments}

E.~J.~C.~D. acknowledges support from the European Union (Marie Sk\l{}odowska-Curie Actions, grant agreement No.~101209876).
The Center for Polariton-driven Light--Matter Interactions (POLIMA) is funded by the Danish National Research Foundation (Project No.~DNRF165).

\section*{Data availability}

All data needed to evaluate the conclusions in the paper are present in the paper and the Supplementary Material. Additional data related to this paper may be requested from the authors.

\section*{Contributions}

J.~D.~C. conceived the study. C.~N.~H. performed the calculations, with input and analysis from all authors. All authors contributed to the development of the theory and writing of the manuscript.

\section*{Conflict of interest}

The authors declare no competing interests.

\end{document}